
\magnification=1200

\def\f#1#2{{\textstyle{#1\over #2}}}

\def\next{\hfil\break\noindent}
\def\R{{\bf R}}
\font\title=cmbx12
\def\next{\hfil\break\noindent}
\hsize=5.5truein
\hfuzz=3.5pt

\noindent
{\title Solutions of the Einstein equations with matter}

\vskip 10pt\noindent
Alan D. Rendall\footnote*
{Present address: Max-Planck-Institut f\"ur
Gravitationsphysik, Schlaatz\-weg 1, 14473 Potsdam, Germany.}
, Institut des Hautes Etudes Scientifiques,
35 Route de Chartres, 91440 Bures-sur-Yvette, France.

\vskip 20pt
\noindent
{\bf Abstract}

\noindent
Recent results on solutions of the Einstein equations with matter
are surveyed and a number of open questions are stated. The first
group of results presented concern asymptotically flat spacetimes,
both stationary and dynamical. Then there is a discussion of
solutions of the equations describing matter in special relativity
and Newtonian gravitational theory and their relevance for
general relativity. Next spatially compact solutions of the
Einstein-matter equations are presented. Finally some remarks
are made on the methods which have been used, and could be used
in the future, to study solutions of the Einstein equations with
matter.

\vskip 20pt
\noindent
{\bf 1. Introduction}

The aim of this paper is to give an overview of recent results on
solutions of the Einstein equations with matter and to present a
number of open questions. The first thing which needs to be done is to
delimit the area to be surveyed. The choice of which subjects are
discussed is of course subjective. The word \lq matter\rq\ indicates,
in the context of this paper, that at least one of the fields which
model the matter content of spacetime in a particular solution can be
physically interpreted as describing matter made up of massive
particles; in other words spacetimes which are empty or which only
contain radiation are not included.  Furthermore, only exact solutions
of the Einstein equations are considered and approximate solutions
(analytic or numerical) are only mentioned insofar as they are
directly relevant to the main topic.  The solutions discussed satisfy
reasonable spatial boundary conditions and are smooth in the sense
that they do not contain any distributional matter sources. The reason
for concentrating on solutions of this type is the wish to consider
situations where the Cauchy problem is known to be well-posed so that
solutions can, if desired, be specified in terms of initial data.
Finally, the phrase \lq exact solutions\rq\ is used here in the literal
sense and not in the sense of \lq explicit solutions\rq, as is common in
general relativity.

There are two main reasons for studying solutions of the Einstein
equations with matter. The first is the possibility of
using them to model astrophysical phenomena. The second is to obtain
insight into questions of principle in general relativity such as the
nature of spacetime singularities and cosmic censorship. Both of these
aspects are discussed in the following.  It should be noted that in
the case of the second of these motivations, the study of the Einstein
equations with matter may even provide insights into the dynamics of
the vacuum Einstein equations. On the one hand there is the
possibility of thinking of a solution of the vacuum equations as
approximately represented by a large scale geometry generated by a \lq
source\rq\ which is the effective energy-momentum of gravitational
waves. On the other hand the inclusion of matter provides a wider range of
examples where it is possible in a relatively simple context of highly
symmetric solutions to gain intuition about the dynamics of the Einstein
equations and develop mathematical tools which can be used to understand
this dynamics. Spherically symmetric spacetimes are the most obvious
examples of the latter possibility.

In order to talk about a solution of the Einstein equations with
matter it is necessary to specify what the matter fields are (i.e.
what kind of geometrical objects are used to describe them), how
the energy-momentum tensor is built from these fields and the
metric and what the matter field equations are. A solution of the Einstein
equations with matter (or a solution of the Einstein-matter equations)
is then a solution of the coupled system of partial differential
equations consisting of the Einstein equations and the matter field
equations. The specific matter models occurring in this paper are the
following:

\noindent
1) perfect fluids described by the Euler equation
\next
2) kinetic theory, where the matter field equation is the
Boltzmann equation (including the special case of collisionless
matter where it reduces to the Vlasov equation)
\next
3) elastic solids
\next
4) any of 1)-3) coupled to an electromagnetic field in an
appropriate manner
\next
All of these matter models may be described as phenomenological.
There is no mention here of matter fields described by fermions
for the simple reason that the author is not aware of any relevant
mathematical results in that case. The subject of viscous fluids is
commented on briefly.

Section 2 presents solutions describing bodies in equilibrium
while section 3 is concerned with dynamical asymptotically flat
solutions. After an interlude on the relations to Newtonian theory
and special relativity in section 4, section 5 describes results
on cosmological solutions. Section 6 contains some reflections on
the methods which have been used to prove theorems about the
Einstein-matter equations up to now.

\vskip .5cm
\noindent
{\bf 2. Solutions describing bodies in equilibrium}

One of the most fundamental types of solutions of the Einstein
equations with matter are those which describe an isolated body
which is at rest or steadily rotating. These are asymptotically
flat and stationary. The simplest case is that of spherically
symmetric static solutions of the Einstein equations with
a perfect fluid as matter model. The existence question for
this case was studied in [1]. Consider a perfect
fluid with energy density $\rho$ and pressure $p$ related by
an equation of state $p=f(\rho)$. Here $f$ is a
continuous non-negative real-valued function on an interval
$[\rho_0,\infty)$, where $\rho_0\ge 0$. Suppose that
$f$ is $C^1$ for $\rho>0$ and $df/d\rho>0$ there. In [1] it was shown
that for each positive real number $p_c$
there exists a unique inextendible spherically symmetric static solution
of the Einstein-Euler equations with a given equation of state and
central pressure $p_c$ which is global in the sense that the area
radius $r$ takes on arbirarily large values. The field equations are in
this situation ordinary differential equations and the theorem
is proved by integrating these equations starting at the
centre. There are two essential points in the proof. The first is
that the equations have a singularity at the centre so that the standard
existence theory for ordinary differential equations is not applicable
there and a replacement must be found. This is provided by a
modification of the contraction mapping argument used to prove
existence and uniqueness in the regular case. The second is that it is
necessary to show that the quantity $2m/r$, where $m$ is the mass function,
remains bounded away from unity, since otherwise the solution could break
down at a finite radius.

This theorem is not the end of the story since there is no
guarantee that the solutions whose existence it asserts are
asymptotically flat; it could be that $m\to\infty$ as $r\to\infty$.
In [1] two cases were exhibited where the solutions
are asymptotically flat. The first is the case $\rho_0>0$. There
it is easy to show that the pressure becomes zero at a finite
radius $r_0$ and an exterior Schwarzschild solution can be matched
there to give a model of a fluid body with finite radius and
mass. Another class of asymptotically flat solutions was obtained
by a rigorous perturbation argument based on the Newtonian
limit. Newtonian models with finite radius for a polytropic fluid were
used as seeds to produce solutions of the Einstein equations with perfect
fluid source using the continuous dependence of the solution of an
ordinary differential equation on parameters. The existence of the
Newtonian solutions is classical [2].

Consider next spherically symmetric static solutions of the Einstein
equations with collisionless matter. Here the matter is thought of as
made up of a cloud of test particles which interact with other only by
the gravitational field which they generate collectively. They are
described statistically by a distribution function $f$ which is the
number density of particles with given position and momentum at a
given time. It satifies the Vlasov (collisionless Boltzmann) equation.
(For information on this equation and general relativistic kinetic
theory in general the reader is referred to [3].)
Each particle travels along a geodesic and as a result of
the assumed symmetries of the spacetime the energy $E$ and modulus of
the angular momentum $F$ of each particle (defined in terms of the
Killing vectors) are conserved. Any distribution function of the form
$f=\phi(E,F)$ for some function $\phi$ is a solution of the Vlasov
equation. This provides a useful method of searching for spherically
symmetric solutions. In [4] the case where $\phi$ is a function of
the energy $E$ alone was studied. This problem is similar to
the fluid problem discussed already and the difficulties which have to
be overcome to obtain an existence and uniqueness theorem are
the same. Once this has been done there remains the question,
for what functions $\phi$ the solutions obtained are asymptotically
flat. A class of solutions with finite radius was obtained in
[4] by perturbing Newtonian solutions, as in the fluid case.
The existence of appropriate Newtonian solutions follows from
the results of [5]. The more general case where
$\phi$ also depends on $F$ was studied by Rein [6]. He
proved the existence of solutions with regular centre and finite
radius as in the simpler case. He also produced solutions
representing a shell of matter surrounding a Schwarzschild black hole.
Interestingly in that case the finiteness of the radius could be
shown directly without using the Newtonian limit. This paper
also contains a method of bounding $2m/r$ which is simpler
and applicable in more general situations than that given in
[1]. (Yet another method can be found in
[7].) There is a conjecture (\lq Jeans' Theorem\rq)
that all static spherically symmetric solutions of the Einstein-Vlasov
equations are such that $f$ depends only on $E$ and $F$. The
corresponding theorem in Newtonian theory has been proved [5].
The solutions of [6] with black holes are counterexamples to
the naive generalization of this statement to general relativity.

\noindent
{\bf Problem} Prove an analogue of Jeans' theorem in general
relativity. Is the naive generalization true in the absence of
black holes?

Under rather general circumstances static solutions of the
Einstein equations with a perfect fluid as matter model are
spherically symmetric. No such general constraint exists
on solutions with collisionless matter. In fact it is probably
the case that the astrophysical objects where a description
by collisionless matter is most accurate are elliptical
galaxies since on the one hand they contain little gas (which
would require a hydrodynamic treatment) and on the other hand they
contain sufficiently many stars to make the statistical model
better than in the case of globular clusters [8].

\noindent
{\bf Problem} Investigate the existence of static solutions of
the Einstein-Vlasov system which are not spherically symmetric.
Are there solutions of this type containing tidally deformed
black holes (i.e. ones not locally isometric to the Schwarzschild
solution)?

\noindent
The equations for spherically symmetric elastic bodies have
been studied in [9] but there is no existence theory
for general constitutive relations known in that case.

\noindent
{\bf Problem} Study the existence of spherically symmetric static
elastic bodies in general relativity

It can be seen from the above that quite a lot is known about
asymptotically flat static solutions of the Einstein-matter equations.
Much less is known about solutions which are stationary but not
static. In fact there is only one existence theorem available in this
case, which is due to Heilig [10].  He proves the existence of rigidly
rotating solutions with compactly supported density for a perfect
fluid having an equation of state belonging to a certain class which
includes equations of state of the form $p=K\rho^\gamma$ with
$1<\gamma<2$.  The method of proof is once again a perturbation
argument, starting fom Newtonian theory. The idea is to start with a
non-rotating Newtonian solution and perturb simultaneously in the
parameter $\lambda=1/c^2$, where $c$ is the speed of light, and the
angular velocity $\omega$ of rotation.  Technically, this is done by
applying the implicit function theorem in a subspace of a weighted
Sobolev space. The subspace is defined by symmetry conditions and is
necessary to get rid of zero eigenvalues of the linearized operator.
This argument is a generalization of the classic work of Lichtenstein
[11] on the existence of rotating fluid bodies in Newtonian theory.
(For a modern treatment of this, see Heilig [12].) The solutions
produced by the perturbation argument are slowly rotating. It would be
interesting to produce more extreme solutions of the kind which can be
observed in numerical computations[13].

\noindent
{\bf Problem} Prove the existence of rapidly rotating solutions
of the Einstein-Euler system.

\noindent
It is also natural to examine possible analogues of Heilig's
theorem for collisionless matter.

\noindent
{\bf Problem} Prove the existence of stationary solutions of
the Einstein-Vlasov system which are not static.

Now we turn to a discussion of results which prove, under certain
hypotheses, that a static solution of
the Einstein-Euler system must be spherically symmetric. The
recent progress on this topic is based on an idea of Masood-ul-Alam
to use the positive mass theorem to prove the spherical symmetry
in a way similar to that in which that theorem had previously been
used to prove uniqueness results for black holes. A given static
solution of the Einstein-Euler system is compared with a reference
solution, which is spherically symmetric and has the same equation
of state and surface potential as the given solution. This idea
was used to prove a theorem in [14] under rather restrictive
conditions on the equation of state. This was improved by Beig
and Simon [15], who extended it to equations of state $p=f(\rho)$
satisfying the inequality
$${\rho+p\over f'}{d\kappa\over d\rho}+2\kappa+{1\over 5}\kappa^2\le 0$$
where $\kappa=(f')^{-1}(\rho+p)/(\rho+3p)$.
However the statement was still only a relative one,
namely that if, for a given solution, a corresponding  reference
solution exists, then the given solution must coincide with the
reference solution and hence be spherically symmetric. The question
of the existence of reference solutions remained open. It was
answered by Lindblom and Masood-ul-Alam [16]. They did not
exactly show the existence of a smooth reference solution but
they did produce a solution of the equations with good enough
properties to make the comparison argument go through. Thus the
proof of spherical symmetry was completed for equations of state
satisfying the above inequality.

\vskip .5cm
\noindent
{\bf 3. Dynamical asymptotically flat solutions}

Once we have solutions describing equilibrium situations, it is
natural to go on to look for dynamical solutions which describe the
motion of bodies, at least on a short time interval. The appropriate
tool for doing this is the Cauchy problem for the Einstein-matter
equations.  While the local Cauchy problem for these equations is well
understood for most types of matter provided there is a strictly
positive lower bound for the density, there are difficulties in
describing bodies, where the density should tend to zero at infinity.
The partial results which are known will now be described. In the
case of a perfect fluid two results are known, both of which only
cover restricted classes of initial data. The first of these is a
theorem of Kind[17,18] in the spherically symmetric
case for equations of state with $\rho_0>0$. The equations are
written in Lagrange coordinates and the method of integration
along characteristics is used. The use of Lagrange coordinates
reduces a problem which a priori involves a (moving) free boundary
to one with a fixed boundary, thus making the problem mathematically
more tractable. Local in time existence and uniqueness of smooth solutions
is shown for initial data with compactly supported density. This
problem is essentially one-dimensional, except for the centre of
symmetry, which has to be handled separately. In higher dimensions
the equations for a perfect fluid described in Lagrange coordinates
are degenerate and it is not known how to handle the initial
value problem directly in this form.

Another way of trying to avoid the difficulties of a moving
boundary is to ignore the boundary and treat the region with fluid
and the exterior vacuum region on the same footing. The problem
with this is the following. In a situation where the density is
bounded below by a positive constant the Euler equations can
be written in symmetric hyperbolic form and this immediately
leads to a local existence and uniqueness theorem. However,
when the equations are put into symmetric hyperbolic form in the
usual way, the hyperbolicity breaks down at any point where the
density vanishes. If the density is everywhere positive but tends
to zero at infinity then the equations are hyperbolic but not
uniformly so, which means that the standard theory does not
guarantee a time of existence of a solution which is uniform
in space. There is a case where the difficulty of the boundary
can be got round. This is a generalization of a trick which
was first used for fluid bodies in Newtonian theory by Makino
[19]. The density is replaced by a new variable $w$ which
is in general a non-smooth function of $\rho$ at the boundary.
After this change of variables the equations can be written in
a form which is symmetric hyperbolic even when the density
vanishes. A local existence theorem for suitable initial data
is an immediate consequence [20,21]. The problem with this
is that \lq suitable\rq\ initial data must be such that $w$
is sufficiently differentiable and this represents a rather
strange restriction when thought of in terms of the energy density
$\rho$, which is the variable with a straightforward physical
interpretation. That this restriction is a serious one can be
seen from the fact [20] that the boundary of the body is
freely falling (i.e. the flow lines of the fluid in the boundary
are geodesic) for all these solutions. Intuitively this means
that the matter is behaving almost like dust near the surface
of the body.

\noindent
{\bf Problem} Prove a local in time existence and uniqueness theorem
for solutions of the Einstein-Euler equations with spatially compact
support and less stringent restrictions on the data than required
for the solutions of Makino type (e. g. small perturbations of
the data coming from a static solution)

\noindent
It is probable that the difficulty of this problem is not just
one of mathematical technique but that there is a physical reason
why it is so delicate. This is the Rayleigh-Taylor instability
which causes fluid interfaces to be violently unstable under
certain circumstances. (For a discussion of this point see
Beale et. al. [22].)

In Newtonian physics it has been shown by Secchi[23] that the Cauchy
problem for the Navier-Stokes equations coupled to gravity is
well-posed locally in time for initial data of compact support
provided the equation of state satisfies $\rho_0>0$ and both
coefficients of viscosity are non-zero. It is well-known that
there are difficulties in describing a relativistic viscous
fluid and, in particular, there does not seem to exist a
preferred relativistic generalization of the Navier-Stokes
equation. (For a recent discussion of relativistic models for
a viscous fluid see [24].) The relativistic models of a viscous
fluid which have been proposed and which have a well-posed Cauchy
problem are hyperbolic, in contrast to the Navier-Stokes equation,
which is parabolic in character. It is thus not clear whether
the incorporation of viscosity in this way leads to an improvement
in the Cauchy problem for a fluid body, as it does in the
Newtonian case.

In kinetic theory there is no qualitative change in the equations in
going from non-vacuum to vacuum and a local existence and uniqueness
theorem for the Einstein-Boltzmann equations with initial density of
compact support is known [25]. For elasticity theory, there are
theorems for the interior of an elastic medium [26] but no existence
theorems for initial data corresponding to an elastic body in general
relativity are known. As in the fluid case it is necessary to face a
moving boundary problem.

\noindent
{\bf Problem} Prove a local existence and uniqueness theorem for an
elastic body in general relativity

Since the only description of matter for which a satisfactory
general local existence theorem for a localized concentration of
matter is known is that given by kinetic theory, it is natural to
start with this kind of matter when looking for global results.
Moreover, within the kinetic description the simplest case is
that of collisionless matter described by the Vlasov equation.
The first result on the global behaviour of asymptotically flat
solutions of the Einstein-Vlasov equations was obtained in [27].
It was shown that the maximal Cauchy development of small,
spherically symmetric initial data for these equations is
geodesically complete. The asymptotic behaviour of the solution
at large times could also be described in some detail. The proof
was carried out using Schwarzschild coordinates. In the
same paper a criterion was obtained for when a solution of
the equations (written in Schwarzschild coordinates) on a given time
interval can be continued to a larger time interval. This criterion
was strengthened in [28] where it was shown that if a solution of
the equations in Schwarzschild coordinates develops a singularity
then the first singularity must be at the centre of symmetry. This
essentially means that the entire problem of showing global
existence of solutions of these equations for large initial data
reduces to controlling the behaviour of the solution near the
centre. This control has not yet been achieved; if it could be
then it should furnish a proof of the weak cosmic censorship
hypothesis for spherically symmetric spacetimes with this matter
model.

Recently, Christodoulou [29] has obtained some interesting
results on the global behaviour of spherically symmetric solutions
of the Einstein-Euler equations in the case of a very special
equation of state. The pressure is zero up to a certain critical
density and for densities above this value the matter is stiff
($dp/d\rho=1$). This equation of state is highly idealized but
can be used as a first approximation in the study of the dynamics
of a supernova. Christodoulou is able to obtain global control
of the solution of the Cauchy problem in this case and he finds
solutions whose global behaviour resembles qualitatively that
of a supernova explosion. There is a collapse followed by the
formation of a shock wave which blows matter off to infinity.
The fact that shock waves are included makes it clear that the
solutions involved are not classical solutions of the Einstein-Euler
equations but weak solutions. However, in contrast to many cases
where the existence of weak solutions of systems of partial
differential equations has been proved, the solutions here are
shown to be uniquely determined by initial data.

The global dynamical solutions mentioned up to now are spherically symmetric.
Unfortunately it seems that once spherical symmetry is abandoned the largest
possible dimension of the isometry group of an asymptotically flat
spacetime without singularities or distributional matter sources is one.
Moreover in this case, which is that of axisymmetry, the single Killing
vector field has fixed points. In our present state of knowledge,
axisymmetric solutions of the Einstein equations do not seem to be
significantly easier to handle than the general case. Note that
there are no results on asymptotically flat solutions of the
Einstein equations with matter approaching in generality the theorem of
Christodoulou and Klainerman [30] concerning the vacuum field equations.
In fact we know nothing about the global dynamics of solutions of the
Einstein-matter equations without symmetry. It should be mentioned that
there is one case which, while not asymptotically flat in the usual sense,
does share some features of asymptotic flatness. This is the case of
cylindrically symmetric spacetimes which are asymptotically flat in all
directions where this is consistent with the symmetry. Solutions of the
source-free Einstein-Maxwell equations with this symmetry have been
studied by Berger, Chru\'sciel and Moncrief[31]. Solutions of the Einstein
equations with matter having this symmetry should also be investigated.

\vskip .5cm
\noindent
{\bf 4. Newtonian theory and special relativity}

To put the global results just mentioned in context, it is
useful to consider the question of solutions of the equations
for matter in special relativity and in the Newtonian theory
of gravity. (One of the advantages of the definition of \lq matter\rq\
used here is that it is typical that each matter model in
general relativity has a reasonable analogue in Newtonian theory.)
It is important to realize that open questions abound even
in these simpler contexts but that there are a number of
interesting new results concerning the Euler equation and the
Boltzmann equation. Some of these will now be reviewed.

Solutions of the classical (compressible) Euler equations typically
develop singularities in finite time [32]. Shock waves are the best
known kind of singularity which occurs. If we wish to study the global
dynamics it is necessary to consider weak solutions. There are various
results known on the existence of global weak solutions in one space
dimension for certain equations of state but there are no uniqueness
theorems in one dimension and no global existence theorems in higher
dimensions. This shows how difficult the study of the compressible Euler
equations is from a mathematical point of view and emphasizes how remarkable
it is that Christodoulou has been able to obtain the results mentioned
in the previous section. Until recently even less was known in the
special relativistic case than in the non-relativistic case. However
Smoller and Temple [33] have now proved a global existence theorem
for the special relativistic Euler equation with a linear equation
of state, so that the difference is no longer so great. Another
question which may be asked is that of the nature of the singularities
which occur when classical solutions of the Euler equation break
down. A partial answer in the case of classical hydrodynamics was
obtained by Chemin[34] who showed that when a classical solution
breaks down the expansion or rotation of the fluid or the gradient
of the density must blow up at some point. Generalizations of this
result to the cases of a self-gravitating fluid in Newtonian theory
and that of a special relativistic fluid have been obtained by
Brauer[35].

In classical hydrodynamics viscous fluids are more tractable than
perfect fluids from a mathematical point of view. In particular, it is
known that given initial data close to equilibrium data there exist
corresponding global smooth solutions of the Navier-Stokes equations.
The proof uses the fact that the Navier-Stokes equations are
essentially parabolic in character. If similar results hold for
relativistic models of viscous fluids then the mechanism which
prevents singularity formation must be more subtle, since the
equations are hyperbolic.

Turning now to kinetic theory, the situation is much better.
Self-gravitating collisionless matter is described in Newtonian
theory by the Vlasov-Poisson system. Many years of study of solutions
of these equations with asymptotically flat initial data culminated
recently in the proof of global existence of classical solutions
for general initial data [36, 37]. The proof can also be adapted
to apply to Newtonian cosmology [38]. Less is known for the
Boltzmann equation but there are nevertheless a number of
significant results. For the classical Boltzmann equation global
existence of classical solutions has been proved for spatially
homogeneous initial data and for data which are small or close to
equilibrium. For general data with finite energy and entropy
global existence of weak solutions (without uniqueness) was
proved by DiPerna and Lions [39]. For information on these results
and on the classical Boltzmann equation in general see
[40,41]. In the last few years some of these results have been extended
to the special relativistic Boltzmann equation. Glassey and Strauss have
proved a global existence theorem for initial data close to equilibrium
[42] while Dudy\'nski and Ekiel-Jezewska[43] have proved a relativistic
analogue of the DiPerna-Lions result on weak solutions. Interestingly
the oldest of the theorems concerning the classical Boltzmann equation
has apparently not yet been extended to the relativistic case, which
leads to the following

\noindent
{\bf Problem} Prove a global in time existence theorem for classical
solutions of the Boltzmann equation in special relativity with spatially
homogeneous intial data or, even better, an appropriate analogue for
spatially homogeneous solutions of the Einstein-Boltzmann equations.

If a particular matter model is such that it develops singularities
when considered as a test field in a given smooth spacetime (and in
particular in special relativity) then it is not surprising if the
system obtained by coupling this matter model to the Einstein
equations develops singularities which have little to do with
gravitational collapse. The best-known example is that of the
shell-crossing and shell-focussing singularities of dust.  It is
simply that the matter variables lose differentiability and cause the
geometry to lose differentiability as well. Hence if matter models of
this kind are permitted, it seems overoptimistic to expect that good
global properties of spacetime such as strong cosmic censorship or the
existence of a global foliation by hypersurfaces of constant mean
curvature will hold. Therefore a good strategy in studying these
problems is to choose only matter models which always have global
smooth solutions corresponding to smooth initial data on a Cauchy
surface in a smooth globally hyperbolic spacetime. Such matter models
will be called tame. A model which is obviously tame is that of
collisionless matter since in that case the field equation in a given
background spacetime is linear. Of course it is easy to think of field
theoretic matter models which are tame because the equations are
linear, e.g. a Maxwell field or a massless scalar field. Evidence that
the restriction to tame matter models is useful in global problems is
provided by the results on crushing singularities[44] which will be
discussed in the next section. The comparison with Newtonian theory
can also be helpful. An example is the insight provided into the
numerical results of Shapiro and Teukolsky [45] by comparing them with
the analytic results for the Vlasov-Poisson system mentioned above (see
[46]). More information on the issue of the choice of matter model in
general relativity can be found in [47].

\vskip .5cm
\noindent
{\bf 5. Spatially compact spacetimes}

Next spatially compact spacetimes will be discussed, i.e. spacetimes
which possess a compact Cauchy surface. These will be referred to in
what follows as cosmological spacetimes since this boundary condition
is appropriate for cosmological models. Of course it is not claimed
that it is the only appropriate one. However the importance of imposing
some spatial boundary condition in mathematical studies of cosmological
models should be emphasized. For without an assumption of this kind
anything can happen; it is impossible to say anything reasonable about the
nature of singularities since any kind of singularity can be built
into the initial data. The assumption of a compact Cauchy surface is
the simplest possibility and therefore a reasonable starting point.
Here the variety of symmetry types which permit the Einstein-matter
equations to be studied under relatively simple conditions is much
richer than in the asymptotically flat case.

The most symmetric cosmological spacetimes are those which are
spatially locally homogeneous. This means by definition that the
universal covering of the spacetime admits a group of isometries with
three-dimensional spacelike orbits. In other word the universal
covering spacetime is spatially homogeneous. The reason for wishing to
include spacetimes which are spatially {\it locally} homogeneous
rather than just those which are spatially homogeneous is that it
allows a much greater variety of symmetry types for a spatially
compact spacetime. If a spatially compact spacetime is spatially
homogeneous then it must have Bianchi I, Bianchi IX or Kantowski-Sachs
symmetry. Generalizing to local homogeneity permits in addition
Bianchi types II, III, V, VI${}_0$, VII${}_0$ and VIII. The
literature on the subject of spatially homogeneous spacetimes is vast
and only a few recent developments will be mentioned here.
In [48] the singularities of spatially compact spatially
locally homogeneous solutions of the Einstein equations were
investigated for a rather general class of phenomenological matter
models. The results were that if the spacetime is not vacuum then for
Bianchi IX and Kantowski-Sachs spacetimes there is a curvature
singularity in both time directions while for the other Bianchi types
there is a curvature singularity in one time direction and the
spacetime is causally geodesically complete in the other time
direction. In particular this proves strong cosmic censorship in this
class of spacetimes. It is interesting to note that in this case the
Einstein equations with matter are easier to analyse than the vacuum
Einstein equations. In the vacuum case the geodesic completeness
statement still holds but the statement about curvature singularities
does not. It fails for spacetimes with Cauchy horizons, such as the
Taub-NUT solution. In the vacuum case it is possible to determine
which spacetimes admit an extension through a Cauchy horizon and
which do not [49,50] but the question of whether those spacetimes
which do not admit extensions through a Cauchy horizon do have
curvature singularities remains open.

While the questions of geodesic completeness and curvature
singularities are of capital importance, it is also of interest to
have more detailed information about the dynamics near the
singularity. In the case where the matter model considered is a
perfect fluid quite a lot is known [51].  On the other hand the
compexity of the dynamics depends significantly on the matter model
chosen. This will be illustrated by two examples.  The first is the
case of solutions of the Einstein-Vlasov system of Bianchi type I.
Note that this is the Bianchi type which is a priori the simplest.
This problem was considered in [52] but only with limited success. The
possible types of dynamical behaviour near the singularity were
reduced to a short list but the question of which types on this list
are actually realized was not answered.  In particular the possibility
was left open that complicated oscillatory behaviour might occur in
this situation. A numerical investigation of the dynamics would be
desirable. With luck it would reveal the answer and this answer could
then be proved to be correct. The other example is due to Leblanc, Kerr
and Wainwright [53] and concerns spacetimes of Bianchi type
VI${}_0$ with perfect fluid and a magnetic field. It was found that
these spacetimes display a behaviour very similar to that of the
Bianchi IX spacetimes with a fluid alone (Mixmaster spacetimes). This
is very interesting for the following reason. A fascinating open
question is whether the complicated Mixmaster behaviour is stable or
whether it would be destroyed by a small inhomogeneous perturbation.
It would be very convenient for numerical or analytical studies if the
Mixmaster spacetime could be perturbed in the class of spacetimes with
two local Killing vectors. Unfortunately any small perturbation of a
Mixmaster spacetime has either no more than one local Killing vector
or it has three, in which case it is again a Mixmaster solution. The
Leblanc-Kerr-Wainwright solutions do not suffer from this problem. A
Bianchi VI${}_0$ spacetime can be perturbed to a spacetime with two
local Killing vectors. These generalized Bianchi VI${}_0$ spacetimes
have Cauchy surfaces whose topology is that of a bundle over a circle
with fibre a torus. The part of the bundle over a small part of the
circle admits a $U(1)\times U(1)$ symmetry group but there is no
corresponding global action. One problem which does remain is that
perturbing a fluid solution is likely to lead to problems with shocks,
so that the matter model used in [53] would need to be replaced by
something else. Nevertheless, this appears a promising approach to
understanding more about the stability of the Mixmaster solution.

Our understanding of the global properties of inhomogeneous
cosmological solutions of the Einstein-matter equations is still very
limited. With one exception, to be mentioned at the end of this section,
theorems are only available in cases with three local Killing vectors.
Note that more is known in the better-studied vacuum case, where there
are results for spacetimes with only two Killing vectors [54]. The
existence of Killing vectors in a spacetime allows the study of
the dynamics of the Einstein equations to be reduced to an effective
problem in lower dimensions. If, however, the dimension of the orbits
is not constant, this leads to singularities in the equations in
the lower dimensional space. In this way a large part of the benefit
of the reduction is lost. It is thus natural to begin by studying
the case where the dimension of the orbits is constant. Spherical
symmetry on $S^2\times S^1$ and plane and hyperbolic symmetry have
this property. In the latter two cases the situation is similar to
that encountered for spatially locally homogeneous spacetimes. If
a compact Cauchy surface is required then the group defining the full
symmetry acts only on the universal cover of spacetime and not on
spacetime itself. Note that it is also possible to have spherically
symmetric spacetimes with spatial topology $S^3$ but there the orbits
are of variable dimension (there are two \lq centres\rq ).
The most precise result on the dynamics of spacetimes
of this type was obtained by Rein [55] who showed that certain kinds
of initial data for the Einstein-Vlasov equations with spherical, plane
or hyperbolic symmetry necessarily lead to curvature singularities where the
Kretschmann scalar $R_{\alpha\beta\gamma\delta}R^{\alpha\beta\gamma\delta}$
blows up uniformly. These data include an open set of all data with the
given symmetry.

Another type of result concerns the question of crushing
singularities.  Recall that a crushing singularity in a cosmological
spacetime is one where a neighbourhood of the singularity can be
covered by a foliation by compact hypersurfaces whose mean curvature
tends uniformly to infinity.  Under mild assumptions on the matter
content of spacetime this is equivalent to the condition that a
neighbourhood of the singularity can be covered by a foliation by
constant mean curvature (CMC) hypersurfaces, whose mean curvature
tends to infinity. The significance of crushing singularities has been
discussed in [56] and [57], where it was suggested that singularities
in cosmological spacetimes should have this property, provided the
matter content of spacetime is sufficiently well-behaved. In [44] and
[58] this was confirmed in the context of solutions of the
Einstein-Vlasov equations with spherical, plane and hyperbolic
symmetry which admit at least one CMC hypersurface. In the spherical
case (on $S^2\times S^1$) the whole spacetime can be covered by a CMC
foliation where the mean curvature takes on all real values. In the
cases of plane and hyperbolic symmetry the CMC foliation obtained was
only shown to cover a neighbourhood of the initial singularity. In a
spacetime with one of these symmetries it is possible to define an \lq
area radius\rq\ as the square root of the area of an orbit. (In the
plane and hyperbolic cases \lq orbit\rq\ should be interpreted as
denoting a subset of spacetime whose inverse image under the
projection from the universal covering space is an orbit.) In Rein's
work [55] this area radius was used as a time coordinate and the
results obtained were stronger than those of [44] and [58]. In fact
the use of $r$ as a time coordinate has a major advantage and a major
disadvantage. The advantage is that this coordinate is optimally
adapted to the symmetry and hence is a powerful tool.  The
disadvantage is that this procedure has no obvious analogue in general
spacetimes, or even spacetimes with less than two Killing vectors. The
CMC approach, on the other hand, is potentially applicable to any
cosmological spacetime. For this reason it would be very desirable to
prove an analogue of Rein's result working directly with the CMC
approach.

Yet another class of results is those which have been obtained by
Burnett [59], [60] on the closed universe recollapse conjecture. He
shows that in a globally hyperbolic spherically symmetric cosmological
spacetime satisfying the dominant energy and non-negative pressures
conditions there is a finite upper bound to the length of all causal
curves. In other words, under these assumptions the universe has a
finite lifetime. In contrast to the other cases mentioned above,
his theorems cover the case of spherical symmetry on $S^3$. It
is nevertheless the case that the proof for $S^3$ is significantly
harder than that for $S^2\times S^1$. This kind of theorem does
not distinguish between spacetime singularities and matter
singularities such as shell-crossing or shocks. However this
is a distinction which it is difficult to make precise in our present
state of knowledge [47].

The most obvious generalization of all these results to a context
with less stringent symmetry assumptions would be to the case of
$U(1)\times U(1)$ symmetry. For the closed universe recollapse
conjecture the spatial topology should be chosen to be $S^3$ or
$S^2\times S^1$, which necessarily leads to a variable dimension
for the group orbits. For the other two types of approach it is
natural to stick to orbits of fixed dimension to start with and
thus to choose the spatial topology $S^1\times S^1\times S^1$. There
is also the possibility of taking spacetimes with some kind of local
$U(1)\times U(1)$ symmetry, such as the generalized Bianchi VI${}_0$
spacetimes mentioned above. These possibilities are now being
investigated in detail. They represent the simplest situations
in which the coupling of localized gravitational waves to matter
(and to each other) can be studied.

The last example of a theorem which gives information about the
global dynamics of solutions of the Einstein equations with
matter which will be discussed here is due to Newman [61]. This is
concerned with the existence of spacetimes with isotropic
singularities, a concept which is related to Penrose's Weyl
curvature hypothesis. The matter model is a perfect fluid with the
equation of state $p=\f13\rho$. It is shown that, without requiring
any symmetry assumptions, it is possible to set up a well-posed
Cauchy problem with initial data given on the singularity itself.
The data which can be given is roughly speaking half the amount which
can be given on a regular spacelike hypersurface and all the spacetimes
produced have isotropic singularities. Since the solution is uniquely
determined by the initial data, if the data are homogeneous and isotropic
then the solution will have Robertson-Walker symmetry.

\vskip .5cm
\noindent
{\bf 6. Methods}

The results discussed in this paper have been obtained by a wide variety
of methods. Some of these have been specially produced to solve particular
problems while others can be seen to be potentially of wider significance.
This section contains some remarks on some of the latter. The
Einstein-matter equations form in general a system of partial differential
equations which is at least in part hyperbolic. The one general technique
available at present for studying hyperbolic equations is the method of
energy estimates. This has been pushed to the limit in the work of
Christodoulou and Klainerman on the vacuum Einstein equations [30]. An
introduction to this method accessible to relativists can be found in
[62]. The method consists in obtaining inequalities for the time
variation of the $L^2$ norm of the unknown quantity $u$ in the
equation, defined by $\|u(t)\|_{L^2}=(\int (u(t,x))^2 dx)^{1/2}$,
and the corresponding norms of derivatives of $u$. In the theory of
nonlinear elliptic equations, which is much more developed than that
of nonlinear hyperbolic equations, it is useful to consider in addition
the $L^p$ norms for $p\ne 2$. Unfortunately it is known that, while
all $L^p$ norms of solutions of elliptic equations have nice properties,
this only holds for hyperbolic equations if $p=2$. There are nevertheless
possibilities of trying to improve on energy estimates by some limited
use of $L^p$ norms. This idea has not yet borne fruit in the case of
the Einstein equations but it has recently been used to obtain a new
result concerning the Yang-Mills equations [63]. This is a local in time
existence and uniqueness theorem for initial data which are only assumed
to have finite energy. Since the energy is conserved, this immediately
implies a global in time existence and uniqueness theorem. In
particular the global existence result of Eardley and Moncrief [64]
is reproduced by a quite new method.

The only one of the global theorems discussed in the above in which
energy estimates play a role is that of Newman. The reason is that in
all the other cases the symmetry assumed is so strong that there are no
locally propagating gravitational waves and correspondingly no truly
hyperbolic phenomena. In Newman's theorem the global result for
the Einstein-matter equations is translated into a local result for
a conformally transformed system. This has some similarity to the
regular conformal field equations used by Friedrich[65] to study the
Einstein equations in vacuum or with certain massless fields. An important
difference is that while Friedrich's equations are regular, Newman's
contain a (mild) singularity, so that it is necessary to do energy
estimates directly. One lower order term in the equations contains
a factor $1/t$, where $t$ is a cosmic time parameter which vanishes
at the singularity and it must be checked that the standard existence
theorem for symmetric hyperbolic systems can be modified to
accomodate this. Wave phenomena appear when the Einstein-matter
equations with $U(1)\times U(1)$ symmetry are considered but even
there, the fact that the hyperbolic equations are effectively only
in one space dimension (at least if the orbit dimension is constant)
means that energy estimates are not the only tool at our disposal.

In a number of the results above an important role is played by
Hawking's quasi-local mass. If $S$ is a compact surface in spacetime
of area $A(S)$ the Hawking mass $m(S)$ is defined to be
$C(A(S))^{1/2}(\chi(S)/2+(1/4\pi)\int_S\rho\mu)$ where $\rho$ and
$\mu$ are the expansions of the two families of future-pointing null
geodesics which start orthogonal to $S$, $\chi(S)$ is the Euler
characteristic of $S$ and $C$ is a constant.  This constant is usually
chosen so that the Hawking mass of a symmetric sphere in the
Schwarzschild solution is equal to the Schwarzschild mass parameter.
However the exact value of the constant is probably of little
significance in general. An alternative expression for the mass is
obtained by applying the Gauss-Bonnet theorem. When this is done the
expression $\int (K+\rho\mu)$ comes up, where $K$ is the Gaussian
curvature.  In order to have a physical interpretation as a mass the
above expression should be non-negative, at least under some
restrictions. It is well known that the Hawking mass is negative for
certain compact surfaces in flat space and this suggests limiting
consideration to surfaces which are in some sense as symmetric as
possible (cf. [66]). In spacetimes with spherical and plane symmetry
surfaces spanned by the local Killing vector fields have Hawking
masses which are non-negative and can only be zero if the spacetime is
flat[44].  The optimism which this might cause is limited by the fact
that in spacetimes with hyperbolic symmetry surfaces spanned by the
local Killing vector fields can have negative mass. In this case the
mass is still of some use for analysing the Einstein equations but
perhaps as a general rule the Hawking mass is most useful in the case
where the surface $S$ is a sphere.

\noindent
{\bf Problem} Find a way to obtain some control of solutions of the
Einstein equations with less than two local Killing vectors using the
Hawking mass (or some variant of it).

\noindent
A technique which seems to be closely related to the Hawking mass
and its positivity was introduced by Malec and \'O Murchadha [67].
They write the constraint equations in a special way in order to
obtain estimates for an initial data set in terms of its mean
curvature. Their technique is a priori limited to spacetimes with
a preferred foliation by surfaces and initial data sets which are
unions of these surfaces. However it may be asked whether it does
not give a hint of some more general property of the Einstein
constraint equations.

The central tool in the proof of Christodoulou and Klainerman of
the nonlinear stability of Minkowski space is the Bel-Robinson
tensor. It is worth noting that Horowitz and Schmidt[68] found, at
least in vacuum, a close connection between the Bel-Robinson
tensor and the Hawking mass of small spheres.

There are various indications that self-similarity emerges in certain
singularities of solutions of the Einstein equations. An illustration
of this is the use of dimensionless variables in the analysis of
Bianchi models [51] which puts the equations in a tractable form,
where self-similar solutions correspond to fixed points of the
dimensionless dynamical system. Another is the privileged position
held by self-similar spherically symmetric spacetimes in the study
of naked singularities in asymptotically flat spacetimes.

\noindent
{\bf Problem} Give a rigorous analysis of spherically symmetric
self-similar solutions of the Einstein-matter equations for a
variety of matter models.

\noindent
The problem to be solved here is to prove the existence of solutions
satisfying reasonable boundary conditions and this in turn comes down
to the qualitative analysis of a system of ordinary differential
equations. A model for an analysis of this type can be found in
the paper [69] of Christodoulou. The question of cutting off a
self-similar solution at a finite radius to get an asymptotically
flat solution is also important.

It would be difficult to make this list of available techniques
which might be used to study general solutions of the Einstein-matter
equations much longer. The conclusion is that all of these should be
studied intensively and that at the same time new techniques should be
sought actively. Therein lies the best hope of the study of the
Einstein-matter equations developing beyond its present embryonic
stage in the coming years. The many new results on global solutions
of partial differential equations obtained recently make it realistic
to expect that a lot of progress can be made in understanding the
global dynamics of solutions of the Einstein equations with matter.

\vskip .5cm
\noindent
{\bf  References}

\noindent
[1] Rendall, A. D.,  Schmidt, B. G.: Existence and properties of spherically
symmetric static fluid bodies with given equation of state.
Class. Quantum Grav. 8, 985-1000 (1991).
\next
[2] Chandrasekhar, S.: An introduction to the study of stellar structure.
Dover, New York, 1957.
\next
[3] Ehlers, J.: Survey of general relativity theory.  In: W. Israel
(ed.) Relativity, Astrophysics and Cosmology. Reidel, Dordrecht,
1973.
\next
[4] Rein, G., Rendall, A. D.: Smooth static solutions of the spherically
symmetric Vlasov-Einstein system. Ann. Inst. H. Poincar\'e (Physique
Th\'eorique) 59, 383-397 (1993).
\next
[5] Batt, J., Faltenbacher, W. and Horst, E.: Stationary spherically
symmetric models in stellar dynamics. Arch. Rat. Mech. Anal. 93,
159-183 (1986).
\next
[6] Rein, G.: Static solutions of the spherically symmetric
Vlasov-Einstein system. Math. Proc. Camb. Phil. Soc. 115, 559-570 (1994).
\next
[7] Baumgarte, T., Rendall, A. D.: Regularity of spherically symmetric
static solutions of the Einstein equations. Class. Quantum Grav.
10, 327-332 (1993).
\next
[8] Binney, J., Tremaine, S.: Galactic dynamics. Princeton University Press,
Princeton, 1987.
\next
[9] Kijowski, J., Magli, G.: A generalization of the relativistic equilibrium
equation for a non-rotating star. Gen. Rel. Grav. 24, 139-158 (1992).
\next
[10] Heilig, U.: On the existence of rotating stars in general relativity.
Commun. Math. Phys. 166, 457-493 (1995).
\next
[11] Lichtenstein, L.: Gleichgewichtsfiguren rotierender
Fl\"ussigkeiten.
\next
Springer, Berlin. 1933.
\next
[12] Heilig, U.: On Lichtenstein's analysis of rotating Newtonian stars.
Ann. Inst. H. Poincar\'e (Physique Th\'eorique) 60, 457-487 (1994).
\next
[13] Herold, H., Neugebauer, G.: Gravitational fields of rapidly
rotating neutron stars: numerical results. In: Ehlers, J., Sch\"afer, G.
(eds.) Relativistic gravity research. Lecture Notes in Physics 410,
Springer, Berlin, 1992.
\next
[14] Masood-ul-Alam, A. K. M.: On spherical symmetry of static perfect
fluid spacetimes and the positive mass theorem. Class. Quantum Grav.
4, 625-633 (1987).
\next
[15] Beig, R., Simon, W.: On the uniqueness of perfect fluid solutions
in general relativity. Commun. Math. Phys. 144, 373-390 (1992).
\next
[16] Lindblom, L., Masood-ul-Alam, A. K. M.: On the spherical symmetry of
static stellar models. Commun. Math. Phys. 162, 123-145 (1994).
\next
[17] Kind, S.: Anfangs-Randwertprobleme f\"ur die Einsteingleichungen
zur Beschreibung von Fl\"ussigkeitskugeln und deren linearisierten
St\"orungen.
\next
Thesis, Munich University.
\next
[18] Kind, S., Ehlers, J.: Initial boundary value problem for the
spherically symmetric Einstein equations for a perfect fluid. Class.
Quantum Grav. 18, 2123-2136 (1993).
\next
[19] Makino, T.: On a local existence theorem for the evolution
equation of gaseous stars. In: Nishida, T., Mimura, M, and Fujii, H.
(eds.) Patterns and Waves. North Holland, Amsterdam, 1986.
\next
[20] Rendall, A. D.: The initial value problem for a class of general
relativistic fluid bodies. J. Math. Phys. 33, 1047-1053 (1992).
\next
[21] Rendall, A. D.: The initial value problem for self-gravitating fluid
bodies. In: K. Schm\"udgen (ed.) Mathematical Physics X. Springer, Berlin,
1992.
\next
[22] Beale, J. T., Hou, T. Y. and Lowengrub, J. S.: Growth rates for
the linearized motion of fluid interfaces away from equilibrium.
Commun. Pure Appl. Math. 46, 1269-1301 (1993).
\next
[23] Secchi, P.: On the equations of viscous gaseous stars. Ann.
Scuola Norm. Sup. Pisa 18, 295-318 (1991).
\next
[24] Geroch, R., Lindblom, L.: Dissipative relativistic fluid theories
of divergence type. Phys. Rev. D41, 1855-1861 (1990).
\next
[25] Bancel, D., Choquet-Bruhat, Y.: Existence, uniqueness and local
stability for the Einstein-Maxwell-Boltzmann system. Commun. Math.
Phys. 33, 83-96 (1973).
\next
[26] Choquet-Bruhat, Y., Lamoureux-Brousse, L.: Sur les \'equations
de
\next
l'\'elasticit\'e relativiste. C. R. Acad. Sci. Paris
276, 1317-1320 (1973).
\next
[27] Rein, G., Rendall, A. D.: Global existence of solutions of the
spherically symmetric Vlasov-Einstein system with small initial data.
Commun. Math. Phys. 150, 561-583 (1992).
\next
[28] Rein, G., Rendall, A. D.  and Schaeffer, J.: A regularity theorem
for solutions of the spherically symmetric Vlasov-Einstein system.
Commun. Math. Phys. 168, 467-478 (1995).
\next
[29] Christodoulou, D.: Self-gravitating fluids: a two-phase model.
Arch. Rat. Mech. Anal. 130, 343-400 (1995) and further papers to appear.
\next
[30] Christodoulou, D., Klainerman, S.: The global nonlinear stability
of the Minkowski space. Princeton University Press, Princeton. 1993.
\next
[31] Berger, B. K., Chru\'sciel, P. T. and Moncrief, V.: On
\lq asymptotically flat\rq\ spacetimes with $G_2$-invariant Cauchy
surfaces. Ann. Phys. 237, 322-354 (1995).
\next
[32] Sideris, T.: Formation of singularities in three-dimensional
compressible fluids. Commun. Math. Phys. 101, 475-485 (1979).
\next
[33] Smoller, J., Temple, B.: Global solutions of the relativistic Euler
equations. Commun. Math. Phys. 156, 65-100 (1993).
\next
[34] Chemin, J.-Y.: Remarques sur l'apparition de singularit\'es dans les
\'ecoulements Euleriens compressibles. Commun. Math. Phys. 133, 323-339
(1990).
\next
[35] Brauer, U.: Unpublished.
\next
[36] Pfaffelmoser, K: Global classical solutions of the Vlasov-Poisson
system in three dimensions for general initial data. J. Diff. Eq. 95,
281-303 (1992).
\next
[37] Lions, P.-L., Perthame, B.: Propagation of moments and regularity
for the three-dimensional Vlasov-Poisson system. Invent. Math. 105,
415-430 (1991).
\next
[38] Rein, G., Rendall, A. D.: Global existence of classical solutions to
the Vlasov-Poisson system in a three dimensional, cosmological setting.
Arch. Rat. Mech. Anal. 126, 183-201 (1994).
\next
[39] DiPerna, R. J., Lions, P.-L.: On the Cauchy problem for Boltzmann
equations: global existence and weak stability. Ann. Math. 130,
321-366 (1989).
\next
[40] Cercignani, C.: The Boltzmann equation and its applications.
Springer, New York, 1988.
\next
[41] Cercignani, C., Illner, R. and Pulvirenti, M.: The mathematical
theory of dilute gases. Springer, New York, 1994.
\next
[42] Glassey, R. T., Strauss, W.: Asymptotic stability of the relativistic
Maxwellian. Publ. Math. RIMS Kyoto 29, 167-233 (1993).
\next
[43] Dudy\'nski, M., Ekiel-Jezewska, M.: Global existence proof for
relativistic Boltzmann equation. J. Stat. Phys. 66, 991-1001 (1992).
\next
[44] Rendall, A. D.: Crushing singularities in spacetimes with spherical,
plane and hyperbolic symmetry. Class. Quantum Grav. 12, 1517-1533
(1995).
\next
[45] Shapiro, S. L.,  Teukolsky, S. A.: Formation of naked singularities:
the violation of cosmic censorship. Phys. Rev. Lett. 66, 994-997 (1991).
\next
[46] Rendall, A. D.: Cosmic censorship and the Vlasov equation. Class.
Quantum Grav. 9, L99-L104 (1992).
\next
[47] Rendall, A. D.: On the choice of matter model in general relativity.
In R. d'Inverno (ed.) Approaches to Numerical Relativity. Cambridge
University Press, Cambridge, 1992.
\next
[48] Rendall, A. D.: Global properties of locally spatially homogeneous
cosmological models with matter. Preprint gr-qc/9409009 (to appear in
Math. Proc. Camb. Phil. Soc.)
\next
[49] Siklos, S. T. C.: Occurrence of whimper singularities. Commun.
Math. Phys. 58, 255-272 (1978).
\next
[50] Chru\'sciel, P. T., Rendall, A. D.: Strong cosmic censorship in vacuum
space-times with compact locally homogeneous Cauchy surfaces. Ann. Phys.
242, 349-385 (1995).
\next
[51] Wainwright, J., Hsu, L.: A dynamical systems approach to
Bianchi cosmologies: orthogonal models of class A. Class. Quantum
Grav. 6, 1409-1431 (1989).
\next
[52] Rendall, A. D.: The initial singularity in solutions of the
Einstein-Vlasov system of Bianchi type I. Preprint IHES/P/95/23,
gr-qc/9505017.
\next
[53] Leblanc, V. G., Kerr, D. and Wainwright, J.: Asymptotic states
of magnetic Bianchi VI${}_0$ cosmologies. Class. Quantum Grav.
12, 513-541 (1995).
\next
[54] Chru\'sciel, P. T., Isenberg, J. and Moncrief, V. Strong cosmic
censorship in polarised Gowdy spacetimes. Class. Quantum Grav.
7, 1671-1680 (1990).
\next
[55] Rein, G.: Cosmological solutions of the Vlasov-Einstein system
with spherical, plane and hyperbolic symmetry. Preprint gr-qc/9409041
(to appear in Math. Proc. Camb. Phil. Soc.)
\next
[56] Eardley, D., Smarr, L.: Time functions in numerical
relativity: marginally bound dust collapse. Phys. Rev. D19,
2239-2259 (1979).
\next
[57] Eardley, D., Moncrief, V.: The global existence problem and cosmic
censorship in general relativity. Gen. Rel. Grav. 13, 887-892 (1981).
\next
[58] Burnett, G., Rendall, A. D.: Existence of maximal hypersurfaces in
some spherically symmetric spacetimes. Preprint gr-qc/9507001.
\next
[59] Burnett, G.: Incompleteness theorems for the spherically symmetric
spacetimes. Phys. Rev. D43, 1143-1149 (1991).
\next
[60] Burnett, G.: Lifetimes of spherically symmetric closed universes.
Phys. Rev. D51, 1621-1631 (1995).
\next
[61] Newman, R. P. A. C.: On the structure of conformal singularities in
classical general relativity. Proc. R. Soc. London A443, 473-492; 493-515
(1993).
\next
[62] Klainerman, S. On the mathematical theory of fields and general
relativity. GR13 Proceedings, IOP, Bristol, 1993.
\next
[63] Klainerman, S., Machedon, M.: Finite energy solutions of the
Yang-Mills equations in $\R^{3+1}$. Ann. Math. 142, 39-119 (1995).
\next
[64] Eardley, D., Moncrief, V.: The global existence of Yang-Mills
fields in $M^{3+1}$. Commun. Math. Phys. 83, 171-212 (1982).
\next
[65] Friedrich, H.: On the global existence and asymptotic behaviour
of solutions to the Einstein-Yang-Mills equations. J. Diff. Geom.
34, 275-345 (1991).
\next
[66] Christodoulou, D., Yau, S.-T.: Some remarks on the quasi-local
mass. Contemporary Mathematics 71, 9-14 (1988).
\next
[67] Malec, E., \'O Murchadha, N.: Optical scalars and singularity
avoidance in spherical spacetimes. Phys. Rev. D50, 6033-6036 (1994).
\next
[68] Horowitz, G. T., Schmidt, B. G.: Note on gravitational energy.
Proc. R. Soc. Lond. A381, 215-224 (1982).
\next
[69] Christodoulou, D.: Examples of naked singularity formation in the
gravitational collapse of a scalar field. Ann. Math. 140, 607-653
(1994).

\end